\documentclass[jkps,twocolumn,fleqn,showpacs,showkeys]{revtex4}
\usepackage{epsf,amsmath,amssymb,verbatim,color,multirow,pifont,mathrsfs,soul,amsthm, wasysym}
\usepackage{graphicx}
\usepackage{bm}
\usepackage{euscript,stmaryrd}
\usepackage{xcolor}
\usepackage[utf8]{inputenc}
\usepackage[T1]{fontenc}
\usepackage{mathptmx}
\begin{document}
\setcounter{page}{0}
\title[]{Discontinuous emergence of a giant cluster in assortative scale-free networks}
\author{Yeonsu Jeong}
\affiliation{Department of Applied Physics, Hanyang University, Ansan 15588, Korea}
\affiliation{Department of Physics, Jeonbuk National University, Jeonju 54896, Korea}
\author{Soo Min Oh}
\affiliation{Center for Theoretical Physics, Seoul National University, Seoul 08826, Korea}
\author{Young Sul Cho}
\email{yscho@jbnu.ac.kr}
\affiliation{Department of Physics, Jeonbuk National University, Jeonju 54896, Korea}
\affiliation{Research Institute of Physics and Chemistry, Jeonbuk National University, Jeonju 54896, Korea}

\today

\begin{abstract}
A giant cluster emerges discontinuously in bond percolation in various networks when the growth of large clusters is globally suppressed.
It was recently revealed that this phenomenon occurs even in a scale-free (SF) network, 
where hubs accelerate the growth of large clusters. The SF network used in the previous study
was disassortative, though, so it is necessary to check whether the phenomenon also occurs in an assortative SF network,
where each hub prefers to be connected to another hub and thus the large cluster growth is accelerated.
In this paper, we find that the phenomenon, namely the discontinuous emergence of a giant cluster in bond percolation with the global suppression of large clusters, also occurs in an assortative SF network.
Interestingly, the generated network is also assortative but not a SF network at the transition point,
unlike the disassortative SF network generated at the transition point in the previous study.
We observe similar behaviors in two additional models and discuss the results.
\end{abstract}


\maketitle

\section{Introduction}
\label{sec:intro}

The explosive percolation (EP) model is a type of percolation model in which a giant cluster emerges abruptly at a transition point
due to the suppression of the growth of large clusters~\cite{jkps_review, jstat_review, nuno_review, souza_nphy}. 
In this model, $N$ isolated nodes are given at the beginning $t = 0$.
At each time step $t \rightarrow t + 1/N$, a link is attached between a disconnected pair of nodes
following a given rule to suppress the growth of large clusters~\cite{Achlioptas:2009, cho_prl:2011, local, yook_etc, yook_etc2, gaussian_prl, bfw_pre},
and the fraction of nodes belonging to the largest cluster denoted by $G$ is the order parameter.
Then, a giant cluster emerges abruptly at a transition point, with $G$ becoming $G > 0$ abruptly as $t$ exceeds the transition point $t_c$.
After the first EP model (the so-called product rule) was introduced~\cite{Achlioptas:2009}, diverse EP models were suggested,
and the transition nature of the models was examined~\cite{friedmanprl:2009, chopre2:2010, filippo_pre:2010, ziff_lattice:2010}. 
As a result, it was concluded that a giant cluster emerges continuously with abnormal critical behaviors when the growth of large clusters is locally suppressed~\cite{mendes_epcont:2010, mendes_epcont:2014, jan_staircase, grassberger_cont, chopre:2021, jan_cont, riordan, cho_science}, whereas a giant cluster emerges discontinuously when the growth of large clusters
is globally suppressed~\cite{jan_cont, riordan, cho_science, smoh:2016, smoh:2018, panagiotou:2011, choprl:2016, sf_hybrid}.

In~\cite{panagiotou:2011}, a rule with an external parameter $g \in (0, 1]$ was applied to the Erd\H{o}s--R\'enyi (ER) model to suppress the growth of large clusters globally.
In this model, two nodes $i, j$ to be connected by a link at each time $t$ are chosen as follows.
\begin{itemize}
\item[(i)] All clusters are ranked in ascending order by size as $\{c_1,c_2,...\}$.
\item[(ii)] A set of nodes $R(t)$ belonging to the $k$ smallest clusters satisfying 
$\sum_{\ell=1}^{k-1}s_{\ell} < \lfloor gN \rfloor \leq \sum_{\ell=1}^k s_{\ell}$ is determined,
where $s_{\ell}$ is the size of the cluster $c_{\ell}$. 
Therefore, $R(t)$ includes approximately $g$ fraction of nodes belonging to the smallest clusters.
\item[(iii)] One node $i$ is selected randomly from the whole set, and another node $j$ is selected randomly from the restricted set $R(t)$.
\end{itemize}
Here, the growth of large clusters is globally suppressed because the connection between a pair of nodes both belonging to the complement of $R(t)$ is not allowed, and global information of the cluster sizes is used in step (i).
This model is called the restricted ER ($r$-ER) model.

The transition nature of the $r$-ER model has been carefully studied in~\cite{panagiotou:2011, choprl:2016}. In these works, it was revealed that the model exhibits a hybrid transition, where a giant cluster emerges discontinuously and
criticality appears at a transition point. In~\cite{sf_hybrid}, it was asked whether the transition nature changes when the global suppression rule
is applied to the generation of a scale-free (SF) network, because the existence of hubs in a SF network would 
accelerate the growth of large clusters, which in turn would compete with the global suppression rule~\cite{sf_hybrid, cho_prl:2009, filippo_prl:2009}. 
To answer this question, the authors considered the static model~\cite{scalefree:2001} to generate SF networks and studied the transition nature when the global suppression rule is applied to the static model.

\begin{figure}[t!]
\includegraphics[width=1.0\linewidth]{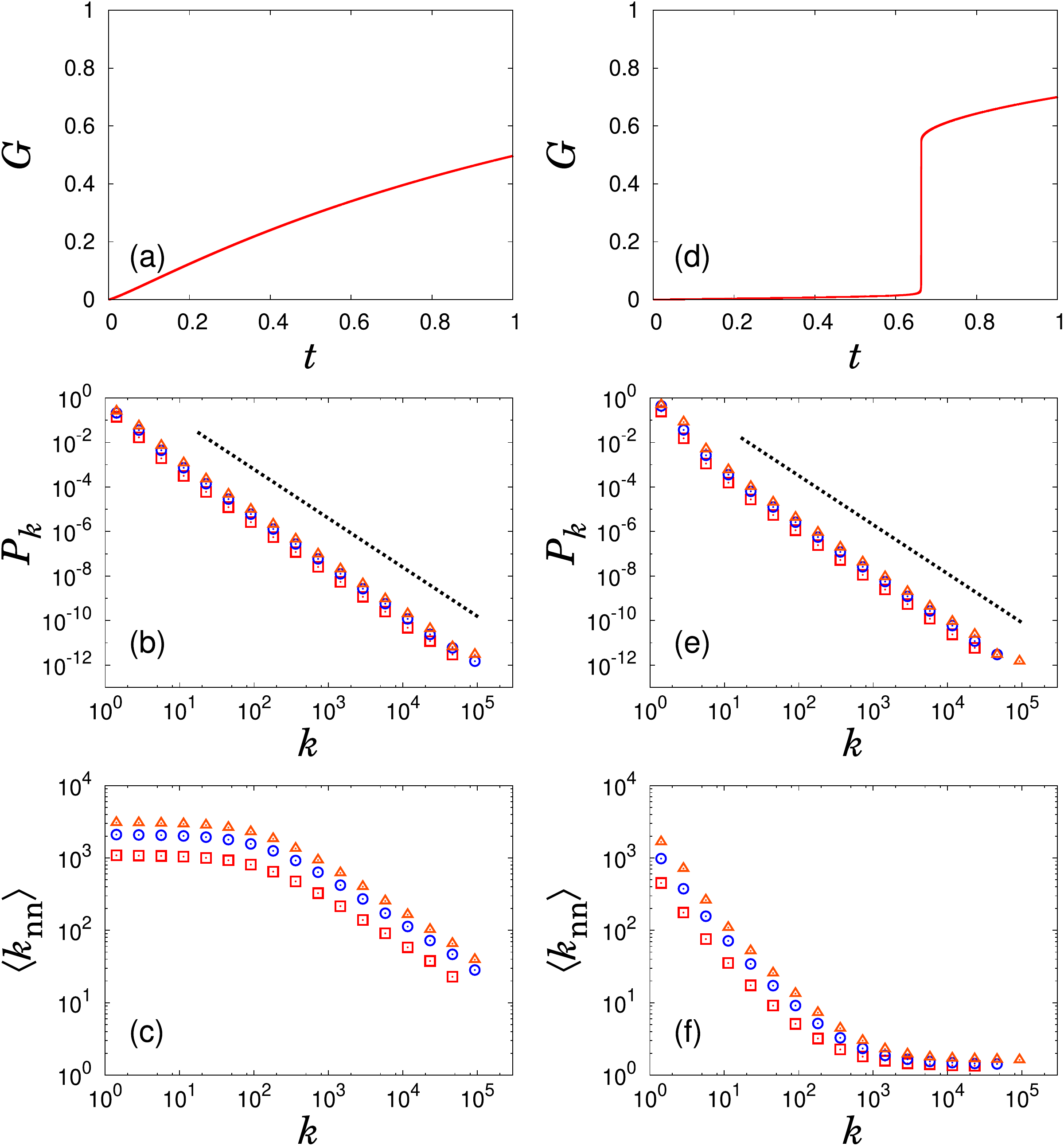}
\caption{(a--c) Simulation data obtained from the static model for $\mu = 5/6$. (a) $G$ vs. $t$ for $N = 2^{10} \times 10^4$.
(b) $P_k$ vs. $k$ for $t=0.2 (\square), 0.4 (\bigcirc),$ and $0.6 (\triangle)$. The dotted line is a guideline with a slope of $-2.2$.
(c) $\langle k_{\text{nn}}\rangle$ vs. $k$ for $t=0.2 (\square), 0.4 (\bigcirc),$ and $0.6 (\triangle)$.
(d--f) Simulation data obtained from the $r$-SF model for $\mu = 5/6$ and $g = 0.5$. (d) $G$ vs. $t$ for $N = 2^{10} \times 10^4$.
(e) $P_k$ vs. $k$ for $t=0.2 (\square), 0.4 (\bigcirc),$ and $0.6 (\triangle)$. The dotted line is a guideline with a slope of $-2.2$.
(f) $\langle k_{\text{nn}}\rangle$ vs. $k$ for $t=0.2 (\square), 0.4 (\bigcirc),$ and $0.6 (\triangle)$.} 
\label{Fig:Rst_G_Pk_knn}
\end{figure}

In the static model, $N$ nodes are indexed $i=1,2,…,N$. 
At each time $t$, two nodes $i, j$ are chosen randomly 
according to the probability $(ij)^{-\mu}/(\sum_{\ell=1}^N{\ell}^{-\mu})^2$. If $i, j$ are not connected yet, they are connected by a link and
$t$ is increased by $1/N$.   
As a result, a SF network with the degree distribution $P_k(k) \propto k^{-\lambda}$ 
is generated, where $\lambda = 1+1/\mu$. It has been shown that a giant cluster emerges continuously from the beginning $t=0$ for $2 < \lambda < 3 (1/2 < \mu < 1)$
in the thermodynamic limit $N \rightarrow \infty$~\cite{cohenpre:2002, scalefree:2004}. These properties are presented in Fig.~\ref{Fig:Rst_G_Pk_knn}(a) and (b). 
We remark that the average nearest neighbor degree of a node of degree $k$ denoted by $\langle k_{\text{nn}}\rangle(k)$ decreases as $k$ increases
in a large $k$ range, as shown in Fig.~\ref{Fig:Rst_G_Pk_knn}(c), and thus the generated SF network is disassortative.
It has also been shown that this degree--degree correlation is an intrinsic property of the static model that arises
because hubs avoid multiple edges and self-loops~\cite{jslee:2006}.

The following process is considered in~\cite{sf_hybrid} to apply the global suppression rule to the static model.
At each time $t$, a set of nodes $R(t)$ is determined with an external parameter $g \in (0, 1]$
following steps (i) and (ii) given above.
Then one node $i$ is selected randomly from the whole set according to the probability $i^{-\mu}/(\sum_{\ell=1}^N{\ell}^{-\mu})$, and 
another node $j$ is selected randomly from the restricted set $R(t)$ according to the probability $j^{-\mu}/(\sum_{\ell \in R}{\ell}^{-\mu})$.
If $i \neq j$ and they are not connected yet, they are connected by a link and $t$ is increased by $1/N$. This model is called the
restricted scale-free ($r$-SF) network model.

The transition nature of the $r$-SF model has been carefully studied in~\cite{sf_hybrid}, where it was revealed that
a hybrid percolation transition is observed as in the $r$-ER model, and that the generated network is a scale-free network.
We check these results in Fig.~\ref{Fig:Rst_G_Pk_knn}(d) and (e): 
$G(t)$ exhibits a shape like that of a hybrid percolation transition in Fig.~\ref{Fig:Rst_G_Pk_knn}(d), and 
SF networks with the degree exponent $\lambda = 1+1/\mu$ are generated in Fig.~\ref{Fig:Rst_G_Pk_knn}(e). 
We remark that $\langle k_{\text{nn}} \rangle (k)$ decreases as $k$ increases, as shown in Fig.~\ref{Fig:Rst_G_Pk_knn}(f), 
and thus the generated SF network is disassortative as in the static model.
Moreover, high-degree nodes have lower values of $\langle k_{\text {nn}} \rangle$ compared to those of the nodes of the static model (Fig.~\ref{Fig:Rst_G_Pk_knn}(c)) with the same $k$ at the same $t$.
We guess that this result may occur because
the global suppression rule in the $r$-SF model induces the high-degree nodes in the complement of $R(t)$ to connect with
low-degree nodes in $R(t)$.

For a fixed degree distribution $P(k) \sim k^{-\lambda}$, it is known that the transition point is advanced 
in bond percolation in an assortative SF network
compared to those in bond percolation in disassortative and uncorrelated SF networks~\cite{dorogovtsev:2008, noh:2007, noh:2008}. 
This is probably because in assortative SF networks, each hub prefers to be connected to another hub and thus the growth of large clusters
is accelerated in bond percolation. 
Therefore, it is still a remaining problem to check whether a giant cluster emerges continuously by overcoming the global
suppression rule when an assortative SF network is used instead of the disassortative SF network used in~\cite{sf_hybrid}.

In this paper, we show that a giant cluster indeed emerges discontinuously at a transition point in bond percolation even in assortative SF networks when the global suppression rule is applied.
Unexpectedly though, we obtain the result that the generated network is not a scale-free network at the transition point.
For simplicity, we do not determine whether the transition is a hybrid transition through further investigation into the existence of criticality at the transition point.

The rest of this paper is organized as follows. In Sec.~\ref{sec:discpt_sf}, we introduce a system
where the global suppression rule is applied to bond percolation in an assortative SF network 
and show that it exhibits a discontinuous emergence of a giant cluster.
We also investigate the degree distribution and degree--degree correlation at the transition point.
In Sec.~\ref{sec:other_models}, we obtain similar results for several variants of the model used in Sec~\ref{sec:discpt_sf}. 
In Sec.~\ref{sec:conclusion}, we discuss the results.

\section{Application of the global suppression rule to bond percolation in assortative SF networks}
\label{sec:discpt_sf}

\begin{figure}[t!]
\includegraphics[width=1.0\linewidth]{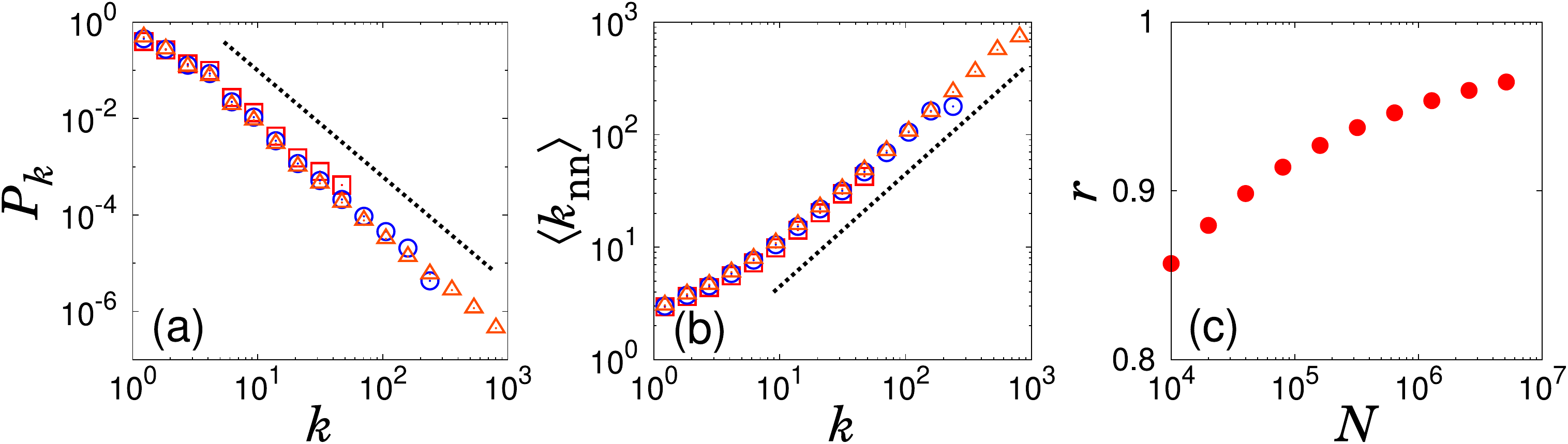}
\caption{Data obtained from assortative SF networks for a fixed $\langle k \rangle = 4$ and $\lambda=2.2$.
(a,b) $P_k$ (a) and $\langle k_{\text{nn}}\rangle$ (b) vs. $k$ for $N/10^4 = 2^0$ $(\square)$, $2^5$ $(\bigcirc)$, and $2^{10}$ $(\triangle)$.
The slopes of the dotted lines in (a) and (b) are $-2.2$ and $1.0$, respectively. (c) $r$ vs. $N$.
} 
\label{Fig:Assorstatic_Pk_knn}
\end{figure}

In this section, we generate assortative SF networks and apply the global suppression rule to bond percolation in them. 
At first, we generate assortative SF networks 
following the rule introduced in~\cite{noh:2008} as described below.
At each time step, one node $i$ ($i=1,...,N$) is chosen randomly 
according to the probability $i^{-\mu}/\sum_{\ell=1}^{N}\ell^{-\mu}$. 
Another node $j$ is selected randomly among the nodes with the same degree as $i$.
If the two nodes $i$ and $j$ are not already connected by a link, a link is attached between them.
We generate a network by attaching $N\langle k \rangle/2$ number of links in this way, where $\langle k \rangle$ is the average degree. 
We note that this process is a variant of the static model modified to connect nodes with the same degree.
It is known that a network generated from this process is an assortative SF network with the degree exponent $\lambda = 1+1/\mu$~\cite{noh:2008}. 
We observe the properties of such generated networks in Fig.~\ref{Fig:Assorstatic_Pk_knn}(a) and (b) where
$P_k \sim k^{-\lambda}$, and $\lambda = 1+1/\mu$ and 
$\langle k_{\text{nn}} \rangle(k) \sim k$
are observed in Fig.~\ref{Fig:Assorstatic_Pk_knn}(a) and (b), respectively.
In Fig.~\ref{Fig:Assorstatic_Pk_knn}(c), we measure the assortativity coefficient $r$, which is the Pearson correlation coefficient between the degrees of neighboring nodes, where $r > 0$ and increases monotonically with $N$.
The results in Fig.~\ref{Fig:Assorstatic_Pk_knn}(b) and (c) support that the generated networks are indeed assortative.

\begin{figure}[t!]
\includegraphics[width=1.0\linewidth]{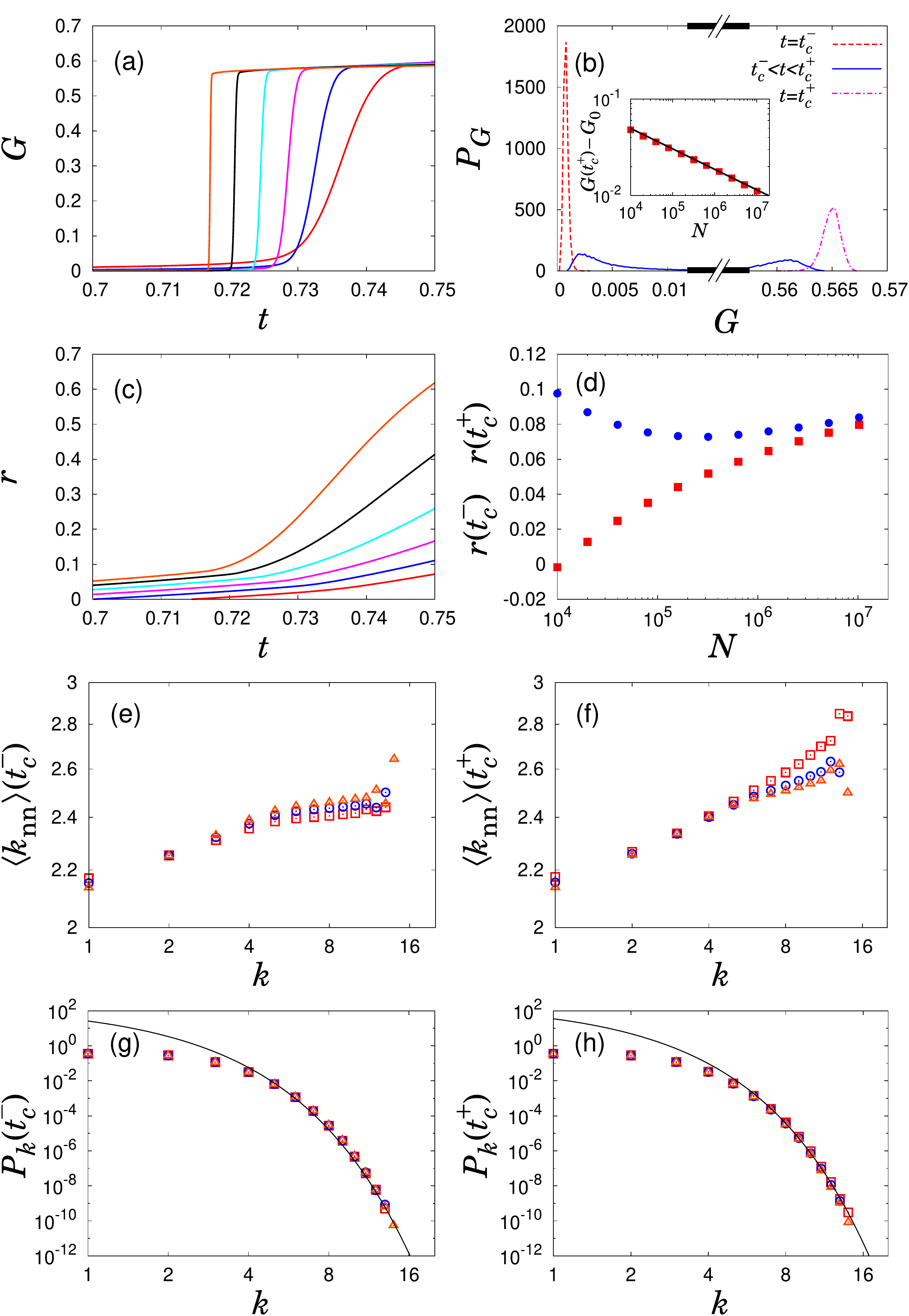}
\caption{Data obtained by applying the global suppression rule to bond percolation in 
the assortative SF networks used in Fig.~\ref{Fig:Assorstatic_Pk_knn}. 
(a) $G$ vs. $t$ for $N/10^4 = 2^0, 2^2, 2^4, 2^6, 2^8, 2^{10}$ from the right.
(b) $P_G$ vs. $G$ at three values of $t$ near the transition point.
Inset: $G(t_c^+)-G_0$ vs. $N$ where $G_0 \approx 0.55$ and slope of the solid line is approximately $-0.21$.
(c) $r$ vs. $t$ for $N/10^4 = 2^0, 2^2, 2^4, 2^6, 2^8, 2^{10}$ from the right. 
(d) $r(t_c^-)$ $(\blacksquare)$ and $r(t_c^+)$ $(\CIRCLE)$ vs. $N$.
(e) $\langle k_{\text{nn}}\rangle(t_c^-)$ for $N/10^4 = 2^6$ $(\square)$, $2^8$ $(\bigcirc)$, $2^{10}$ $(\triangle)$.
(f) $\langle k_{\text{nn}}\rangle(t_c^+)$ for $N/10^4 = 2^6$ $(\square)$, $2^8$ $(\bigcirc)$, $2^{10}$ $(\triangle)$.
(g) $P_k(t_c^-)$ vs. $k$ for $N/10^4 = 2^6$ $(\square)$, $2^8$ $(\bigcirc)$, $2^{10}$ $(\triangle)$. 
(h) $P_k(t_c^+)$ vs. $k$ for $N/10^4 = 2^6$ $(\square)$, $2^8$ $(\bigcirc)$, $2^{10}$ $(\triangle)$.
The solid lines in (g) and (h) are exponential functions proportional to $\text{exp}(-ax)$ with $a \approx 2.0$.} 
\label{Fig:Rstlattice_G_Pk_knn}
\end{figure}

We occupy the bonds in the generated networks where the nodes are connected by occupied bonds.
We now apply the global suppression rule 
to bond percolation in the generated networks as described below. 
At each time $t$, the set $R(t)$ of approximately $gN$ nodes
included in the smallest clusters is determined with an external parameter $g \in (0, 1]$
following steps (i) and (ii) in Sec.~\ref{sec:intro}.
An unoccupied bond is then randomly selected and occupied unless both nodes of the bond
belong to the complement of $R(t)$. Figure~\ref{Fig:Rstlattice_G_Pk_knn} presents the simulation results of this model.

In Fig.~\ref{Fig:Rstlattice_G_Pk_knn}(a), $G$ increases more abruptly near the transition point as $N$ increases.
In Fig.~\ref{Fig:Rstlattice_G_Pk_knn}(b), we follow the analysis in~\cite{sf_hybrid}
to show that $G$ exhibits a discontinuity in the thermodynamic limit $N \rightarrow \infty$.
Specifically, we measure distributions of $G$ denoted by $P_G$ by changing $t$ near the transition point for a fixed $N$. 
From this, we find that there exist $t_c^-$ and $t_c^+$, where $P_G$ has double peaks for $t_c^- < t < t_c^+$ and
that the peak in the small (large) $G$ region disappears as $t$ becomes larger (smaller) than $t_c^+$ $(t_c^-)$.
Therefore, a giant cluster with the size $G(t_c^+)$ similar to the location of the peak of $P_G$ in the large $G$ region 
emerges as $t$ exceeds $t_c^+$. 
In this respect, $G(t_c^+)$ can be regarded as the giant cluster size just after the transition. 
As shown in the inset of Fig.~\ref{Fig:Rstlattice_G_Pk_knn}(b), $G(t_c^+)$ converges to $G_0 > 0$ as $G(t_c^+)-G_0 \sim N^{-0.21}$.
We also check that $t_c^+ - t_c^- \rightarrow 0$ and $G(t_c^-) \rightarrow 0$ as $N \rightarrow \infty$,
such that $G$ indeed exhibits a discontinuity with the finite gap size $G_0 > 0$ at $t_c$,
where $t_c$ is the limit of $t_c^+, t_c^-$ as $N \rightarrow \infty$.

In Fig.~\ref{Fig:Rstlattice_G_Pk_knn}(c), $r(t)$ drastically increases for $t > t_c^+$. 
To understand this phenomenon,
we observe $G(t_c^+) > (1-g)$, which results in all nodes belonging to $R(t)$ thereby rendering the global suppression rule ineffective for $t > t_c^+$.
Consequently, this model reduces to random bond percolation in the assortative SF networks,
and $r(t)$ drastically increases to the
$r$ values in Fig.~\ref{Fig:Assorstatic_Pk_knn}(c) as $t$ increases.

In Fig.~\ref{Fig:Rstlattice_G_Pk_knn}(d), both $r(t_c^+), r(t_c^-)$ are positive and increase with $N$
such that we expect $r(t_c)>0$ in the limit $N \rightarrow \infty$. This result means that an assortative network is constructed at $t_c$
in the assortative SF networks 
even though the global suppression rule induces the constructed network to be disassortative, as
described in Sec.~\ref{sec:intro} when discussing the difference between Fig.~\ref{Fig:Rst_G_Pk_knn}(c) and (f).
As expected from $r(t_c^+), r(t_c^-) > 0$ in Fig.~\ref{Fig:Rstlattice_G_Pk_knn}(d),
$\langle k_{\text{nn}}\rangle(t_c^-)$ and $\langle k_{\text{nn}}\rangle(t_c^+)$ increase with $k$
as shown in Fig.~\ref{Fig:Rstlattice_G_Pk_knn}(e) and (f), respectively.

Interestingly, $P_k(t_c^-)$ and $P_k(t_c^+)$ decrease exponentially as shown in Fig.~\ref{Fig:Rstlattice_G_Pk_knn}(g) and (h), respectively,
even though the underlying network follows $P_k \sim k^{-\lambda}$. In other words, the constructed networks at $t_c$
are not SF networks, in contrast to the underlying network being a SF network.  
This result also differs from the construction of SF networks at the transition point when the global suppression rule 
is applied to the static model, see Fig.~\ref{Fig:Rst_G_Pk_knn}(e)~\cite{sf_hybrid}.

We interpret this phenomenon qualitatively as follows. 
If the global suppression rule is applied to bond percolation
in the SF networks, hubs may belong to the complement of $R$. Since the bond between nodes belonging to the complement of $R$ cannot be occupied according to the global suppression rule,
link occupation between hubs may not be allowed. 
The underlying networks in our model are assortative SF networks,
whereas the static model generates disassortative SF networks.
In the assortative SF networks of our model, the neighbors of each hub are mostly hubs, 
as shown in Fig.~\ref{Fig:Assorstatic_Pk_knn}(b), unlike the disassortative SF networks generated by the static model. 
Therefore, most of the bonds connected to each hub 
are not occupied up to $t_c$ where the global suppression rule is effective in the assortative SF networks of our model. 
As a result, $P_k$ of the occupied bonds at $t_c$ decreases exponentially,
as shown in Fig.~\ref{Fig:Rstlattice_G_Pk_knn}(g) and (h).

\begin{figure}[t!]
\includegraphics[width=1.0\linewidth]{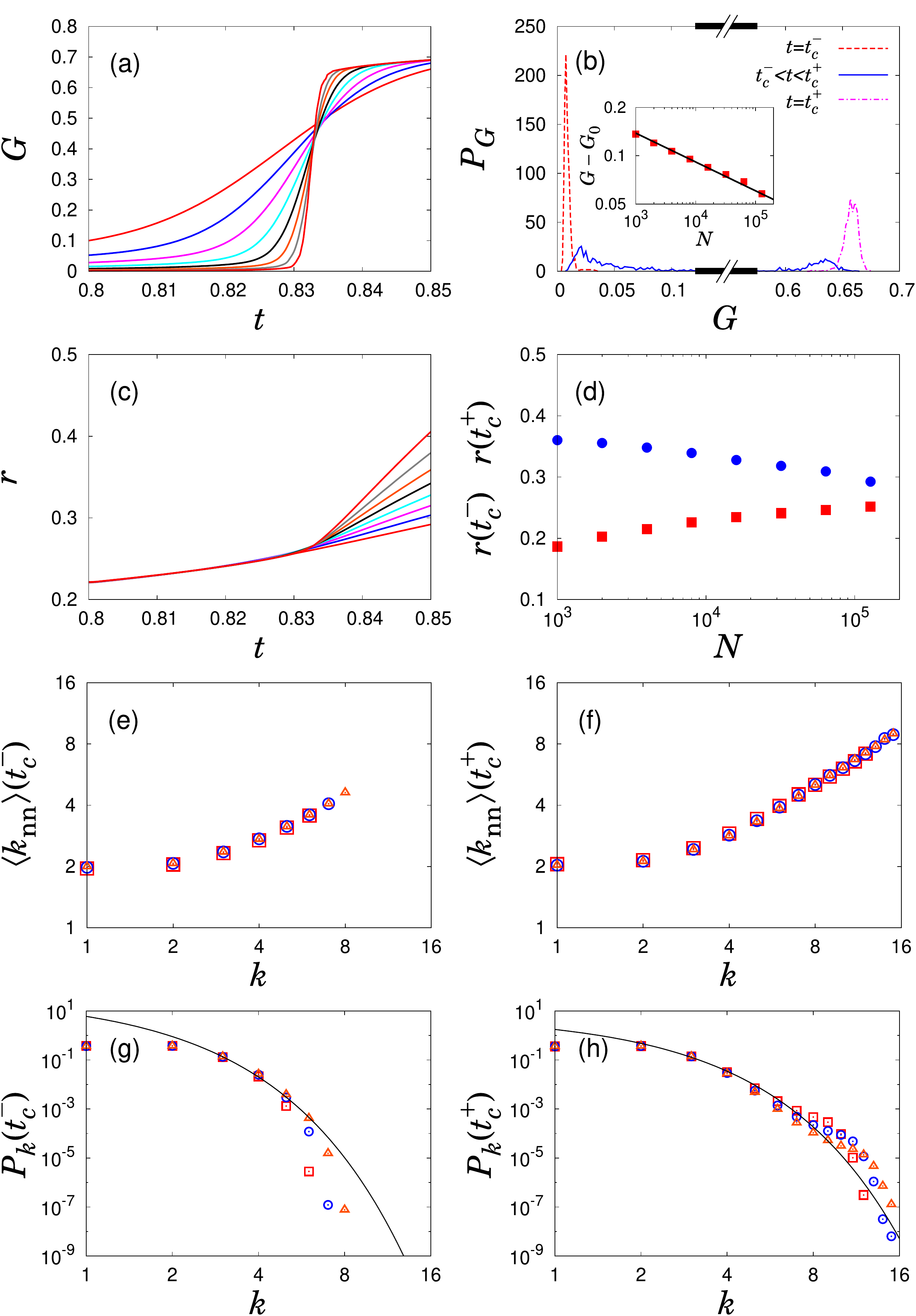}
\caption{Data are obtained from the modified $r$-SF model with $\lambda=2.2$ and $g=0.5$,
where a node in the set $R$ is connected to the node with the most similar degree. 
(a) $G$ vs. $t$ for $N/10^3 = 2^0, 2^1, 2^2, 2^3, 2^4, 2^5, 2^6, 2^7$, where the maximum slope increases as $N$ increases.
(b) $P_G$ vs. $G$ at three values of $t$ near the transition point.
Inset: $G(t_c^+)-G_0$ vs. $N$ where $G_0 = 0.6$ and the slope of the solid line is $-0.18$.
(c) $r$ vs. $t$ for $N/10^3 = 2^0, 2^1, 2^2, 2^3, 2^4, 2^5, 2^6, 2^7$ from the bottom up. 
(d) $r(t_c^-)$ $(\blacksquare)$ and $r(t_c^+)$ $(\CIRCLE)$ vs. $N$.
(e) $\langle k_{\text{nn}}\rangle(t_c^-)$ for $N/10^3 = 2^3$ $(\square)$, $2^5$ $(\bigcirc)$, $2^7$ $(\triangle)$.
(f) $\langle k_{\text{nn}}\rangle(t_c^+)$ for $N/10^3 = 2^3$ $(\square)$, $2^5$ $(\bigcirc)$, $2^7$ $(\triangle)$.
(g) $P_k(t_c^-)$ vs. $k$ for $N/10^3 = 2^3$ $(\square)$, $2^5$ $(\bigcirc)$, $2^7$ $(\triangle)$. 
(h) $P_k(t_c^+)$ vs. $k$ for $N/10^3 = 2^3$ $(\square)$, $2^5$ $(\bigcirc)$, $2^7$ $(\triangle)$.
The solid lines in (g) and (h) are exponential functions proportional to $\text{exp}(-ax)$ with $a \approx 1.9$ and $a \approx 1.3$, respectively.} 
\label{Fig:RstB_G_Pk_knn}
\end{figure}

\begin{figure}[t!]
\includegraphics[width=1.0\linewidth]{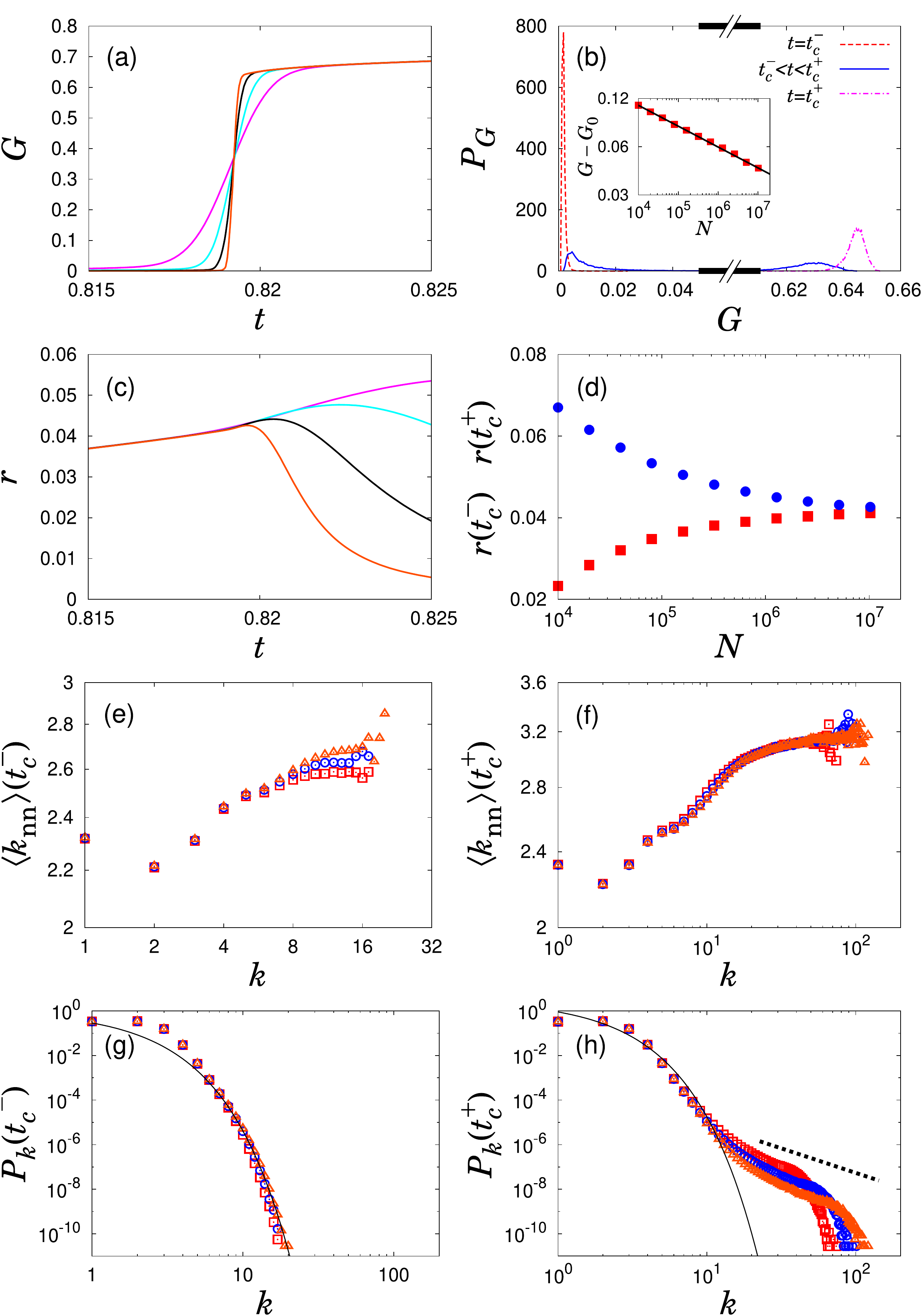}
\caption{Data are obtained from the modified $r$-SF model with $\lambda=2.2$ and $g=0.5$,
where a node in the set $R$ is connected to the node with a smaller degree difference between two node candidates.
(a) $G$ vs. $t$ for $N/10^4 = 2^4, 2^6, 2^8, 2^{10}$, where the maximum slope increases as $N$ increases.
(b) $P_G$ vs. $G$ at three values of $t$ near the transition point.
Inset: $G(t_c^+)-G_0$ vs. $N$ where $G_0 = 0.6$ and the slope of the solid line is $-0.13$.
(c) $r$ vs. $t$ for $N/10^4 = 2^4, 2^6, 2^8, 2^{10}$ from the top down. 
(d) $r(t_c^-)$ $(\blacksquare)$ and $r(t_c^+)$ $(\CIRCLE)$ vs. $N$.
(e) $\langle k_{\text{nn}}\rangle(t_c^-)$ for $N/10^4 = 2^6$ $(\square)$, $2^8$ $(\bigcirc)$, $2^{10}$ $(\triangle)$.
(f) $\langle k_{\text{nn}}\rangle(t_c^+)$ for $N/10^4 = 2^6$ $(\square)$, $2^8$ $(\bigcirc)$, $2^{10}$ $(\triangle)$.
(g) $P_k(t_c^-)$ vs. $k$ for $N/10^4 = 2^6$ $(\square)$, $2^8$ $(\bigcirc)$, $2^{10}$ $(\triangle)$. 
(h) $P_k(t_c^+)$ vs. $k$ for $N/10^4 = 2^6$ $(\square)$, $2^8$ $(\bigcirc)$, $2^{10}$ $(\triangle)$, where the
slope of the dotted line is $-2.2$.
The solid lines in (g) and (h) are exponential functions proportional to $\text{exp}(-ax)$ with $a \approx 1.2$.} 
\label{Fig:RstBm2_G_Pk_knn}
\end{figure}

\section{Modified restricted scale-free network models to reduce the degree difference between neighbors}
\label{sec:other_models}

In the previous section, we applied the global suppression rule to bond percolation in assortative SF networks.
In the model, a giant cluster emerges discontinuously, and the constructed network is an assortative network
with a positive assortativity coefficient at the transition point. Moreover, it was shown that the constructed network is not a SF network even though the underlying network is a SF network.

In this section, we introduce two additional models and show that they exhibit similar behaviors.
In both models, at each time $t$, the set $R(t)$ of approximately $gN$ nodes is determined
for the external parameter $g \in (0, 1]$ following steps (i) and (ii) in Sec.~\ref{sec:intro},
and a node $i \in R(t)$ is chosen randomly according to the probability $i^{-\mu}/(\sum_{\ell \in R}\ell^{-\mu})$.
The difference between the two models appears when $i$ selects another node to connect to.

In the first additional model, $i$ connects to the node with the most similar degree among all nodes. 
In the second additional model, two candidates $j_1$ and $j_2$ are randomly selected among all nodes, and then $i$ connects to the node with a smaller difference in degree between the two candidates.

We note that these models are variants of the $r$-SF model~\cite{sf_hybrid} modified to reduce the difference in degree between neighbors. 
Moreover, we choose $i$ from the restricted set $R(t)$ rather than from the entire system, for the following reason. 
If $i$ is selected randomly from the entire system according to the probability 
$i^{-\mu}/(\sum_{\ell=1}^N\ell^{-\mu})$ while the node to be connected with $i$ is selected from the restricted set $R(t)$, 
$i$ will be mainly a hub. Accordingly, the degree difference between these nodes should be large because $R(t)$ may contain only low-degree nodes and thus no hubs.

Figure~\ref{Fig:RstB_G_Pk_knn} presents the results of applying the analysis used in Fig.~\ref{Fig:Rstlattice_G_Pk_knn} 
to the first additional model. In Fig.~\ref{Fig:RstB_G_Pk_knn}(a), $G$ increases more abruptly near the transition point as $N$ increases.
In Fig.~\ref{Fig:RstB_G_Pk_knn}(b), we follow the analysis in Fig.~\ref{Fig:Rstlattice_G_Pk_knn}(b) 
and show that $G$ exhibits a discontinuity in the thermodynamic limit $N \rightarrow \infty$
with the finite gap size $G_0 > 0$ at $t_c$, where $G_0$ and $t_c$ are limits of $G(t_c^+)$ and $t_c^+$ $(t_c^-)$ as $N \rightarrow \infty$, respectively.
In Fig.~\ref{Fig:RstB_G_Pk_knn}(c), $r$ increases drastically for $t > t_c^+$. To understand this phenomenon,
we observe $G(t_c^+) > (1-g)$, where all nodes belong to $R(t)$.
Therefore, links would mainly connect between hubs, and thus $r$ would increase drastically for $t > t_c^+$.  
In Fig.~\ref{Fig:RstB_G_Pk_knn}(d), we expect $r(t_c)>0$ in the limit $N \rightarrow \infty$. 
This result means that an assortative network is constructed at $t_c$
because of the rule that $i \in R$ is connected to the node with the most similar degree among all nodes,
even though the global suppression rule induces the constructed network to be disassortative,
as described in the discussion of the difference between Fig.~\ref{Fig:Rst_G_Pk_knn}(c) and (f) in Sec.~\ref{sec:intro}.
In Fig.~\ref{Fig:RstB_G_Pk_knn}(e) and (f), $\langle k_{nn} \rangle$ increases with $k$ near $t_c$
as expected through Fig.~\ref{Fig:RstB_G_Pk_knn}(d). In Fig.~\ref{Fig:RstB_G_Pk_knn}(g), $P_k(t_c^-)$ decreases faster than exponential. 
This result shows that link attachments to hubs are effectively inhibited up to $t_c^-$ because of the rule that $i \in R$ is connected to the node with the most similar degree among all nodes.
In Fig.~\ref{Fig:RstB_G_Pk_knn}(h), $P_k(t_c^+)$ decreases exponentially with a bump in the large $k$ region.
We interpret that the bump forms because link attachments between hubs occur dominantly up to $t=t_c^+$ after $G$ exceeds $1-g$.

In Fig.~\ref{Fig:RstBm2_G_Pk_knn}, similar results for the second additional model are presented. 
In (a) and (b) in the figure, it is shown that $G$ exhibits a discontinuity at $t_c$ with the gap size $G_0>0$ in the thermodynamic limit $N \rightarrow \infty$.
In Fig.~\ref{Fig:RstBm2_G_Pk_knn}(c) and (d), $r(t_c) > 0$ and thus an assortative network is constructed at $t_c$. 
However, $r(t)$ decreases as $t$ increases beyond $t_c^+$, as shown in Fig.~\ref{Fig:RstBm2_G_Pk_knn}(c),
unlike the behaviors of the other models presented in Fig.~\ref{Fig:Rstlattice_G_Pk_knn}(c) and Fig.~\ref{Fig:RstB_G_Pk_knn}(c).
We check the case where $G(t_c^+)>1-g$ and thus all nodes belong to $R(t)$ for $t>t_c^+$. 
In other words, two candidate node pairs, $(i, j_1)$ and $(i, j_2)$,
are chosen randomly among all nodes according to the probabilities $(ij_1)^{-\mu}/(\sum_{\ell=1}^N \ell^{-\mu})^2$ and 
$(ij_2)^{-\mu}/(\sum_{\ell=1}^N \ell^{-\mu})^2$, respectively, and the pair with the smaller degree difference is connected for $t > t_c^+$.
We note that disassortative SF networks are generated in the static model, 
where a pair of nodes $(i,j)$ is chosen randomly according to the probability $(ij)^{-\mu}/(\sum_{\ell=1}^N \ell^{-\mu})^2$ and connected.
Therefore, we guess that the competitive rule using only two link candidates would generate disassortative networks, and thus $r$ decreases as $t$ increases beyond $t_c^+$.
In Fig.~\ref{Fig:RstBm2_G_Pk_knn}(e) and (f), $\langle k_{nn} \rangle$ increases with $k$ near $t_c$
as expected through Fig.~\ref{Fig:RstBm2_G_Pk_knn}(d).
In Fig.~\ref{Fig:RstBm2_G_Pk_knn}(g), $P_k(t_c^-)$ decreases exponentially and SF networks are not generated
because link attachments to hubs are effectively inhibited by the competitive rule up to $t_c^-$.
In Fig.~\ref{Fig:RstBm2_G_Pk_knn}(h), $P_k(t_c^+)$ shows a short power-law regime in the large $k$ region,
where the regime forms because link attachments between hubs occur dominantly up to $t=t_c^+$ after $G$ exceeds $1-g$.

\section{Conclusion}
\label{sec:conclusion}

In summary, we modified rules to make the degrees of connected nodes become similar
in hybrid percolation transition models generating disassortative SF networks at the transition point.
In the three models that we used, assortative networks are generated at the transition points as expected, but the generated networks are not SF networks.
We also checked that a giant cluster emerges discontinuously at the transition point.

The motivation of this work was to observe continuous transitions by accelerating connections between hubs
even when the global suppression rule is applied.
However, continuous transitions were not observed in the models that we considered.
In~\cite{noh:2007}, it was reported that percolation in assortative networks share common properties with
percolation in growing networks. Therefore, the existing result in~\cite{smoh:2018} that a giant cluster emerges discontinuously
when the global suppression rule is applied to percolation in growing networks might explain our results.
Furthermore, applying the global suppression rule to percolation in assortative growing networks can be an attempt to
observe continuous transition by overcoming the global suppression rule.

We note that all of these models fail to generate assortative SF networks at the onset of the transitions.
Therefore, future work should find a new rule that generates assortative SF networks at the onset of the transition
when applied to hybrid percolation transition models. 
Studying the transition nature of such a model would be of interest.

\section*{Acknowledgement}
This work was supported by a National Research Foundation of Korea (NRF) grant, No. 2020R1F1A1061326.

\end{document}